\documentclass[a4paper,aps,pra,showpacs,twocolumn,superscriptaddress]{revtex4-2}

\usepackage[utf8]{inputenc}  %Input what you want e.g., é, ł, a, ü
\usepackage[T1]{fontenc}     %Output what you want e.g., é, ł, a, ü
\usepackage[british]{babel}  %Do hyphenation according to british english
\usepackage{lmodern}  %Joe's font

\usepackage[scaled=1.03]{inconsolata} %Monospace font
\usepackage[usenames,dvipsnames]{color} %More color choices
\usepackage[colorlinks,citecolor=blue,linkcolor=magenta,urlcolor=blue]{hyperref}  %Hyperlinks (pink, green, blue)
\usepackage{graphicx} % Package to insert external figures
\usepackage{tikz}
\usepackage[babel]{microtype}  %Improves text justification
\usepackage{amsmath,amssymb,amsthm,bm,mathtools,amsfonts,mathrsfs,bbm,dsfont} %Useful math packages
\usepackage{xspace}  %Useful to add space in macros
\usepackage{pgf}
\usepackage{newtxtext}
\usepackage{newtxmath}
\usepackage[cal=cm]{mathalfa}

\usepackage{multirow}
\usepackage{verbatim}
\usepackage{enumerate}
\usepackage{tcolorbox}
\usepackage[normalem]{ulem}
\usepackage{orcidlink}
\usepackage{hyperref}
\usepackage{siunitx}

\usepackage{physics}
\usepackage[mathscr]{eucal}

\usepackage[dvipsnames]{xcolor}
\definecolor{memory}{HTML}{d948bc}

\usepackage{tikz,quantikz}
\usepackage{tikz}
\usetikzlibrary{patterns}
\usetikzlibrary{patterns.meta}

\definecolor{mathematicablue}{rgb}{0.87,0.94,1}
\definecolor{mathematicadarkblue}{rgb}{0.368417, 0.506779, 0.709798}
\definecolor{mathematicaorange}{rgb}{1,0.9,0.8}
\definecolor{mathematicadarkorange}{rgb}{1,0.5,0}

\newcommand{\id}{\ensuremath{\vvmathbb{1}}}

\newcommand{\Bcal}{\mathcal{B}}
\newcommand{\Ecal}{\mathcal{E}}

\newcommand{\Tcal}{\mathcal{T}}

\newcommand{\Lcal}{\mathcal{L}}
\newcommand{\Ccal}{\mathcal{C}}
\newcommand{\Scal}{\mathcal{S}}

\newcommand{\Kcal}{\mathcal{K}}
\newcommand{\Dcal}{\mathcal{D}}
\newcommand{\Gcal}{\mathcal{G}}

\newcommand{\s}{\vec{s}}
\newcommand{\wit}{W_3}
\newcommand{\BBS}{W_2}

\newcommand{\cpt}{\Ecal}

\makeatletter
\def\maketitle{
\@author@finish
\title@column\titleblock@produce
\suppressfloats[t]}
\makeatother

\definecolor{darkorange}{RGB}{255,100,0}

\begin{document}

\title{Observation of Genuine Multi-Time Quantum Memory}

\author{Konstantin Beyer\,\orcidlink{0000-0002-1864-4520}}
\affiliation{Department of Physics, Stevens Institute of Technology, Hoboken, New Jersey 07030, USA}
\email{kbeyer1@stevens.edu}

\author{Valentin Link\,\orcidlink{0000-0002-1520-7931}}
\affiliation{Institut für Physik und Astronomie, Technische Universität Berlin, D-10623, Berlin, Germany}
\email{valentin.link@tu-berlin.de}

\author{Surihan Sean Borjigin\,\orcidlink{0000-0002-4325-150X}}
\affiliation{Department of Electrical Engineering and Computer Science, University of California Berkeley, Berkeley, CA, USA, 94720.}

\author{Qian Cao\,\orcidlink{0009-0004-2848-2280}}
\affiliation{Department of Electrical Engineering and Computer Science, University of California Berkeley, Berkeley, CA, USA, 94720.}

\author{Orion Lee\,\orcidlink{0000-0002-9860-398X}},
\affiliation{Department of Electrical Engineering and Computer Science, University of California Berkeley, Berkeley, CA, USA, 94720.}

\author{Kater Murch\,\orcidlink{0000-0002-1536-294X}}
\affiliation{Department of Electrical Engineering and Computer Science, University of California Berkeley, Berkeley, CA, USA, 94720.} 

\author{Igor Pikovski\,\orcidlink{0000-0002-9441-2553}}
\affiliation{Department of Physics, Stevens Institute of Technology, Hoboken, New Jersey 07030, USA}
\affiliation{Department of Physics, Stockholm University, SE-106 91 Stockholm, Sweden}

\author{Philip Taranto\,\orcidlink{0000-0002-4247-3901}}
\affiliation{Department of Physics \& Astronomy, University of Manchester, Manchester M13 9PL, United Kingdom}
\email{philip.taranto@manchester.ac.uk}

\date{\today}

\begin{abstract}
Non-Markovian dynamics in open quantum systems arise from memory effects in the environment. Crucially, this memory can often be explained classically and does not require the storage of quantum information. The presence of genuine quantum memory can be certified by existing criteria that compare the evolution of the system at two different times. We show that quantum memory can remain hidden from all such pairwise tests and become detectable only when three or more times are considered jointly. We derive an analytical three-time witness and use it to certify quantum memory in a minimal qubit example in which every pair of times has a classical-memory explanation, thus establishing a strict hierarchy of quantum memory effects. Genuine multi-time quantum memory does not rely on fine-tuned time-dependent Hamiltonians or gate-based evolution. We experimentally demonstrate the hierarchy of quantum memory in a time-continuous realization on a superconducting-qubit platform with time-independent couplings. 
\end{abstract}

\maketitle

\textit{Introduction.---}Quantum systems that interact with their environment are highly sensitive to its properties, allowing the dynamics of open quantum systems to reveal features of the environment that are never directly observed~\cite{haikkaNonMarkovianQuantumProbes2014,zwickMaximizingInformationEnvironment2016,tamascelliQuantumProbingPure2020,plodzienHarnessingSpinqubitDecoherence2025}. A prominent example is non-Markovianity, where information exchanged with the environment can later influence the system's evolution and give rise to memory effects~\cite{breuerColloquiumNonMarkovianDynamics2016,Devega_2017,shrikantQuantumNonMarkovianityOverview2023,gaikwadEntanglementAssistedProbe2024,bernardesExperimentalObservationWeak2015a}. Crucially, however, non-Markovianity does not certify \textit{quantum} memory, since the same dynamics may admit a realization based solely on \textit{classical} memory~\cite{giarmatziWitnessingQuantumMemory2021,backer_local_2024,smirneConnectionMicroscopicDescription2021,banackiInformationBackflowMay2023,taranto_2024,rossetResourceTheoryQuantum2018,buscemiCausalNoncausalRevivals2025,milzWhenNonMarkovianQuantum2020,santosTwopointMeasurementCorrelations2025}. To infer a genuinely quantum memory resource, one must instead show that reproducing the observed dynamics necessarily requires storing coherent quantum information in the environment. The system then serves as a probe of an intrinsic quantum property of its surroundings~\cite{beyerOnesidedWitnessQuantumness2025}.
Beyond this fundamental aspect, the distinction between classical and quantum memory has direct implications for the complexity of simulating open quantum systems and many-body dynamics~\cite{banulsMatrixProductStates2009a,sonnerInfluenceFunctionalManybody2021,leroseInfluenceMatrixApproach2021,folignoTemporalEntanglementChaotic2023,vilkoviskiy2026Temporal}, since classical memory can be sampled stochastically~\cite{kahlertSimulatingLandauZener2024,wang2026Temporal}.

A practical route to this inference is offered by snapshots of the system's dynamical map. These are completely positive and trace-preserving (CPTP) maps (quantum channels) $\cpt_{n}$ that describe the evolution from a common initial time $t_0=0$ to later times \(t_n\), obtained by standard channel tomography. 
Recent work has shown how to certify quantum memory by ruling out classical-memory explanations for channels observed at two different times~\cite{backer_local_2024}, e.g., from the pair $(\cpt_{1},\cpt_{2})$. Such an explanation would mean that, at the earlier of the two times (e.g., $t_1$), all information that the environment retains about the system can be reduced to a classical measurement record, which is then fed back to condition the subsequent evolution (from $t_1$ to $t_2$). All existing criteria operate in this pairwise manner, testing the different channels two at a time~\cite{backerEntropicWitnessQuantum2025a,yu_quantum_2025,beyerOnesidedWitnessQuantumness2025,backerVerifyingQuantumMemory2026,backerRevealingQuantumNature2026}. This raises the question whether the quantum nature of memory can always be decided at the level of such pairwise tests.

Here, we show that this is \textit{not} the case. We derive an analytical three-time witness that certifies quantum memory using three channels of a qubit dynamics at once. We then construct a qubit dynamics for which every pair of channels admits a classical-memory realization, yet the three-time witness implies that no single classical-memory dynamics can reproduce the full triple jointly. The classical-memory models that can realize different pairs of channels are thus mutually incompatible and cannot be combined into a consistent description of the full dynamics. 
This establishes a hierarchy in temporal quantum-memory correlations in dynamical maps, analogous to the distinction between pairwise and genuinely multipartite quantum correlations~\cite{miklinMultiparticleEntanglementEmergent2016}. Quantum memory can be hidden from every pairwise test while remaining detectable through genuinely multi-time observations. 

The fact that quantum memory effects can have multi-time structure, irreducible to correlations between pairs of times, has previously been established at the level of process tensors \cite{Milz_2017,Pollock_2018_PRL,Pollock_2018_PRA,Taranto_2019_PRL,Taranto_2019_PRA,Taranto_Thesis,Taranto_2022,milzGenuineMultipartiteEntanglement2021}, and observed in experiments that intervene on the system at intermediate
times~\cite{guoExperimentalDemonstrationInstrumentspecific2021,white2020Demonstration}. However, such direct observation requires control
interventions during the dynamics and an experimental effort that scales exponentially in the number of time steps~\cite{Milz_2018,white2022NonMarkovian}. In contrast, our compatibility witness instead accesses a genuinely multi-time property of the dynamics through the dynamical map alone. This is what makes the observed effect experimentally accessible. No interventions are needed throughout the evolution, and the experimental effort scales only linearly in the number of times probed. 

We demonstrate this accessibility by realizing a continuous-time open quantum system dynamics with transmon qubits on a superconducting chip. Channel tomography of a single qubit at different times is sufficient to apply our three-time quantum memory criterion, while explicit classical-memory models---constructed numerically for every pair of measured channels---confirm that no pairwise test could reveal it. A multi-time analysis thus yields a strictly stronger signature for quantum memory based on the same tomographic data. Together, these results identify genuine multi-time quantum memory as a distinct and experimentally accessible resource in open quantum dynamics. 

%%%%%%%%%%%%%%%%%%%%%%%%%%%%%%%%%%%%%%%%%%%%%%

\textit{Dynamics with Quantum Memory.---}A dynamical map \(\cpt_t\) on the time domain \(\Tcal\) exposes \emph{two-time} quantum memory if there is at least one pair of times \(t_1 < t_2 \in \Tcal\) for which the two maps \(\cpt_1\) and \(\cpt_2\) (where \(\cpt_n := \cpt_{t_n}\)) \emph{cannot} be explained using classical memory alone. More precisely, one asks whether the pair \((\cpt_1, \cpt_2)\) admits a decomposition of the form
\begin{align}
\label{eq:biclassical-decomposition}
    \cpt_1[\rho] = \sum_i K_i \rho K_i^\dagger, && \cpt_2[\rho] = \sum_i \Phi_i\left[ K_i \rho K_i^\dagger \right],
\end{align}
where the Kraus operators \(K_i\) satisfy \(\sum_i K_i^\dagger K_i = \id\), and the \(\{\Phi_i\}\) are CPTP maps. If no such decomposition exists, the pair \((\cpt_1, \cpt_2)\) certifies genuinely quantum memory~\cite{backer_local_2024}.

Operationally, Eq.~\eqref{eq:biclassical-decomposition} describes a scenario in which \textit{all} information retained by the environment at the intermediate time $t_1$ is reduced to a classical label $i$. The Kraus operators $K_i$ define the possible branches of this classical description, while the subsequent evolution $\Phi_i$ may depend on the value of $i$. This is precisely a measurement-and-feedback scenario: The environment acts as a measuring apparatus that stores the outcome $i$ in a classical register, and the subsequent evolution is conditioned on that outcome. Information about the past can thus influence the future dynamics, but only through a classical measurement record. In other words, no coherent quantum information is carried forward by the environment.
Existing witnesses of quantum memory based on the dynamical map rely on this two-time definition: They determine whether each pair of observed channels 
admits a common classical-memory explanation~\cite{backer_local_2024,yu_quantum_2025,backerEntropicWitnessQuantum2025a,beyerOnesidedWitnessQuantumness2025,backerVerifyingQuantumMemory2026}.

Similarly, a dynamical map \(\cpt_t\) exhibits \emph{three-time} quantum memory effects if there exists a triple \(t_1 < t_2 < t_3 \in  \Tcal\) such that the corresponding channels \(\cpt_1\), \(\cpt_2\) and $\cpt_3$ \emph{cannot} be realized through 
\begin{gather}
    \cpt_1[\rho] = \sum_i K_i \rho K_i^\dagger, \quad \cpt_2[\rho] = \sum_{ij} G_{j|i} K_i \rho K_i^\dagger G_{j|i}^\dagger, \notag \\
    \cpt_3[\rho] = \sum_{ij} \Phi_{ij}\left[  G_{j|i} K_i \rho K_i^\dagger G_{j|i}^\dagger\right],
    \label{eq:triclassical-decomposition}
\end{gather}
where \(\sum_j G_{j|i}^\dagger G_{j|i} = \id \, \forall \, i\) and \(\{\Phi_{ij} \}\) are CPTP maps. 
This decomposition likewise admits a classical-memory interpretation. The Kraus operators $K_i$ generate an initial classical record \(i\) in the environment, while the operators $G_{j|i}$ describe a subsequent update of that record. The condition $\sum_j G_{j|i}^\dagger G_{j|i} = \id \, \forall \, i$ ensures that this update defines a valid conditional quantum channel. The pair of labels $(i,j)$ therefore represents a classical memory that updates as the dynamics progresses, and the final channel $\Phi_{ij}$ may depend on the complete classical history. At every stage, information is propagated through the environment only via classical measurement records and conditional feedback, with no coherent quantum memory transmitted between successive observation times.

More generally, one can define $k$-time quantum memory effects by allowing $k$ ordered classical memory updates. Any dynamics exhibiting $k$-time quantum memory effects necessarily exhibits $k'$-time quantum memory for all $k' \geq k$. This raises the question whether this hierarchy is strict: Can quantum memory remain hidden from every pairwise test while still becoming detectable when three observation times are considered jointly? 
In the following, we answer this question in the affirmative, demonstrating both theoretically and experimentally that dynamics where quantum memory can be revealed only through joint consideration of several observation times exist.
Note that, in contrast, for the weaker question of whether the dynamics requires memory at all, pairwise tests are conclusive since memoryless pairs of maps can be composed directly into a \(k\)-step memoryless process~\cite{rivasQuantumNonMarkovianityCharacterization2014}. 

\textit{Witnessing Three-Time Quantum Memory.---}The preceding discussion suggests a natural compatibility problem. Given three channels $(\cpt_1,\cpt_2,\cpt_3)$, does there exist a single three-time classical-memory decomposition of the form in Eq.~\eqref{eq:triclassical-decomposition} reproducing all of them simultaneously? 
We here present a witness that rules out such a decomposition for a given triple of qubit channels.

Let \(\cpt_1,\cpt_2,\cpt_3\) be qubit CPTP maps. We first define the \textit{displacement vector}
\begin{align}
\label{eq:s-displacement}
    \s_1 := \tr\big[\cpt_1[\id/{2}] \Vec{\sigma} \big], \quad \Vec{\sigma} = (\sigma_x,\sigma_y,\sigma_z)^\top.
\end{align}
This vector describes the image of the maximally mixed state under $\cpt_1$, and its magnitude quantifies the degree of non-unitality of the channel.

For a unit vector $\hat{n}$, let
\begin{align}
\label{eq:antipodal-states}
    \rho_\pm^{\hat n} := \tfrac{1}{2}\left( \id \pm \hat n\!\cdot\!\Vec{\sigma}\right) = \ketbra{\pm_{\hat n}}{\pm_{\hat n}}
\end{align}
denote the pair of antipodal pure states on the Bloch sphere. The fidelity between these states after the application of the channel \(\cpt_k\) is then defined by
\begin{align}
\label{eq:Fk}
    F_k(\hat n) := F\!\left(\cpt_k[\rho_+^{\hat n}], \cpt_k[\rho_-^{\hat n}]\right), 
\end{align}
where \(F(\rho, \chi) := \tr[\sqrt{\sqrt \rho \chi \sqrt \rho}]\) is the \textit{quantum fidelity}.

For every triple of channels admitting a decomposition of the form in Eq.~\eqref{eq:triclassical-decomposition}, the quantity
\begin{align}
    \label{eq:triquantum-witness-definition}
        \wit := \min_{\hat{n},\hat{m}} \left[F_2(\hat n) + F_3(\hat m) - \abs{\s_1}  \, \abs{\hat n \cross \hat m} \right]
\end{align}
satisfies
\begin{align} \label{eq:triquantum-witness-condition}
    \wit \geq 0.
\end{align}
Consequently, an observed violation $\wit < 0$ rules out \textit{every} possible three-time classical-memory realization of the form in Eq.~\eqref{eq:triclassical-decomposition} and hence \(\wit\) is a witness that certifies quantum memory.

The intuition behind the witness is an information--disturbance
trade-off. A nonzero displacement vector \(\vec{s}_1\) implies that the Kraus operators \(\{K_i\}\) that generate the classical record at \(t_1\) [see Eq.~\eqref{eq:biclassical-decomposition}] necessarily
realize a measurement that extracts information about the system (the operators cannot be unitary). Each
outcome of such a measurement leaves antipodal states orthogonal along
at most one direction and irreversibly reduces their distinguishability
along every other direction. If antipodal states nevertheless remain highly distinguishable at later times along two incompatible directions $\hat{n}$ and $\hat{m}$, then no classical record \(i\) can account for the observed dynamics. The proof is given in App.~\ref{app:proof}. 

Although a violation of Eq.~\eqref{eq:triquantum-witness-condition} certifies quantum memory, it does not by itself imply that the quantum memory is hidden from pairwise tests. Indeed, the violation could arise because one of the pairs $(\cpt_1,\cpt_2)$, $(\cpt_1,\cpt_3)$, or $(\cpt_2,\cpt_3)$ already lacks a two-time classical-memory explanation. To establish the stronger notion of \textit{genuine} multi-time quantum memory, one must additionally show that each pair admits a decomposition of the form in Eq.~\eqref{eq:biclassical-decomposition}. In that case, the observed quantum memory \textit{cannot} be attributed to any pairwise incompatibility, but arises solely from the impossibility of combining the pairwise classical explanations into a single consistent description of the full three-time dynamics. 

\textit{A Minimal Example.---}Before we show the hierarchy of quantum memory effects for a time-continuous dynamical map, we illustrate the concept of genuine three-time quantum memory with an analytical minimal discrete example: three qubit channels \((\cpt_1, \cpt_2, \cpt_3)\) that are pairwise decomposable with classical memory as in Eq.~\eqref{eq:biclassical-decomposition} but that cannot be obtained with three-time classical memory as in Eq.~\eqref{eq:triclassical-decomposition} when the full triple is considered.

\begin{figure}[t]
    \centering
    \includegraphics[width=\linewidth]{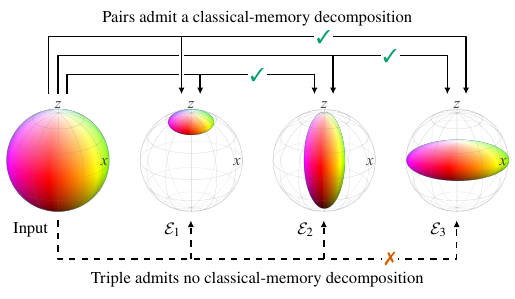}
    \caption{Bloch sphere visualization of a discrete qubit dynamics that exhibits genuine three-time quantum memory effects. Despite the evolution being strongly non-Markovian, each pair of maps \((\cpt_1,\cpt_2)\), \((\cpt_1,\cpt_3)\), and \((\cpt_2,\cpt_3)\) (solid lines) can be realized with classical memory as in Eq.~\eqref{eq:biclassical-decomposition}. The triple \((\cpt_1,\cpt_2,\cpt_3)\) (dashed lines) does not admit a classical-memory realization of the form in Eq.~\eqref{eq:triclassical-decomposition}.    
    We note that the ellipsoid representing \(\cpt_3\) cannot be reached from the ellipsoid of \(\cpt_2\) by a unitary rotation (as indicated by the color coding).}
    \label{fig:discrete_bloch}
\end{figure}

Our proposed channels are visualized as Bloch sphere images in Fig.~\ref{fig:discrete_bloch}.
The first map \(\cpt_1\) is an amplitude damping channel with strength \(\gamma\), the maps \(\cpt_2\) and \(\cpt_3\) are dephasing channels of strength \(p\) around the \(z\) and \(x\) axis, respectively (see App.~\ref{sm::discreteexample} for definitions).
The pair \((\cpt_2, \cpt_3)\) can be decomposed with classical memory as in Eq.~\eqref{eq:biclassical-decomposition} since \(\cpt_2\) is of mixed-unitary form~\cite{landauBirkhoffTheoremDoubly1993,backerQuantumMemoryPrecludes2026a}.
The pairs \((\cpt_1, \cpt_2)\) and \((\cpt_1,\cpt_3)\) can be shown to have a classical-memory decomposition iff \(p \geq p_\text{crit} := (1-\sqrt{1-\gamma})/2\) (see App.~\ref{sm::biclassical-decompositions}).
However, the triple shows \(\wit<0\) for any damping \(\gamma > 0\). Choosing \(\hat{n} = (0,0,1)^\top =: \hat{z}\) and \(\hat m = (1,0,0)^\top =:\hat{x}\)  [see Eq.~\eqref{eq:triquantum-witness-definition}], we find \(F_2(\hat z) = F_3(\hat x) = 0\) since the dephasing maps \(\cpt_{2,3}\) preserve the orthogonality of the \(\sigma_z\) and \(\sigma_x\) eigenstates, respectively [see also Eq.~\eqref{eq:Fk}].
On the other hand, \(\cpt_1\) is non-unital for \(\gamma > 0\) and therefore \(\abs{\s_1}>0\). Hence, \(\wit <0\) in Eq.~\eqref{eq:triquantum-witness-definition} and the triple \((\cpt_1,\cpt_2,\cpt_3)\) exhibits genuine three-time quantum memory effects for any \(p\geq p_\text{crit}\) (where the pairwise classical decomposition exists).

\textit{Time-continuous Dynamics.---}The dynamics constructed above establishes the existence of three-time quantum memory effects that can remain hidden from every pairwise test. 
We now turn to a time-continuous dynamical map on an interval \(\Tcal = (0,T)\) that exhibits the same phenomenon and implement it experimentally. Our model is based on a simple Markovian embedding in terms of a time-independent Hamiltonian, which shows that this effect appears as a generic phenomenon and does not require fine-tuning, time-dependent control, or gate-based simulation. 

We consider a probe qubit $S$ coupled to two independent qubit environments $A$ and $B$ through Hamiltonians
\begin{gather}
\label{eq:Hamiltonian-SA}
    H_{SA} =  \hbar g \,(\sigma_+^S\otimes \sigma_-^A+\sigma_-^S \otimes \sigma_+^A), \\
    \label{eq:Hamiltonian-SB}
    H_{SB} = \hbar \omega \,\sigma_x^S \otimes \sigma_x^B.
\end{gather}
Preparing both environments in their ground states $\ket{0}$, the interaction $H_{SA}$ produces the familiar non-Markovian amplitude-damping dynamics with excitation revivals, while $H_{SB}$ generates coherent dephasing and rephasing about the $x$-axis. Together, they produce a non-Markovian dynamical map on $S$
\begin{align}
\label{eq:dynamical-map-with-B}
    \cpt_t[\rho_S] = \tr_{AB}[U_t(\rho_S \otimes \ketbra{0}{0}_A \otimes \ketbra{0}{0}_B)U_t^\dagger],
\end{align}
where \(U_t = \exp[-i t(H_{SA}+H_{SB})/\hbar]\).

Although the model as stated involves two environmental qubits, its physical mechanism is particularly transparent. Since $H_{SB}$ couples through $\sigma_x^B$, the environmental qubit $B$ effectively acts as a coherent control determining the sign of a local $x$-rotation on the probe system \(S\). From the perspective of the reduced dynamics of $S$, the model is therefore equivalent to a statistical mixture of two driven $SA$ evolutions
\begin{align}
    \cpt_t[\rho_S] = \frac{1}{2} \sum_{i=\pm} \tr_{A}[U^{(i)}_t(\rho_S \otimes \ketbra{0}{0}_A)U_t^{(i)\dagger}], 
    \label{eq:model-dynamics}
\end{align}
with \(U_t^{(\pm)} := \exp[-i t(H_{SA} \pm \hbar\omega \sigma_x^S)/\hbar]\). 
In this form, the model can experimentally be implemented on a two-qubit platform.

\textit{Experimental Realization.---}Our experimental implementation uses two fixed-frequency transmon qubits on a superconducting circuit platform~\cite{Koch2007,Qian2026}. The probe qubit $S$ and memory qubit $A$ have transition frequencies $\omega_S/(2\pi)=4.080\,\si{GHz}$ and $\omega_A/(2\pi)=4.000\,\si{GHz}$, respectively. The two qubits are coupled through a flux-tunable transmon coupler. Each qubit has an individual microwave control line for rotations in the $x$--$y$ plane and is dispersively coupled to a readout resonator~\cite{Blais2004, Walter2017}. Parametric modulation of the coupler activates the exchange interaction $H_{SA}$ in Eq.~\eqref{eq:Hamiltonian-SA}~\cite{Caldwell2018,Didier2018}. We tune the modulation amplitude to set $g/(2\pi)=0.994\,\si{MHz}$. The microwave drive line on $S$ implements the local term $\pm\hbar\omega\sigma_x^S$. The microwave drive amplitude therefore sets the ratio $\omega/g$ (see Supplemental Material (SM)~\ref{sm::experimental-calibration} for details). 

To characterize the channel in Eq.~\eqref{eq:model-dynamics}, we prepare different initial states, evolve for variable durations, and then measure the qubits.  We prepare the probe in the six cardinal states
$\ket{0},\ket{1},\ket{\pm x},\ket{\pm y}$ and the memory in the state $\ket{0}$. We then apply the exchange interaction
and local drive simultaneously.  Finally, we  measure the probe in the three Pauli
bases. For every evolution time, we acquire the two signs of the local drive and
average the resulting channels, realizing the statistical mixture in
Eq.~\eqref{eq:model-dynamics}. We use the measurement results to reconstruct the channel $\cpt_t$. The
quantities $\abs{\s(t)}$, $F_t(\hat z)$, and $F_t(\hat x)$ are then
calculated directly from these reconstructed channels [via Eqns.~(\ref{eq:s-displacement},\ref{eq:Fk})] and combined according
to Eq.~\eqref{eq:triquantum-witness-definition} to form the witness \(\wit\). 

\begin{figure}[t]
    \centering
    \includegraphics[width=\linewidth, trim={0 1cm 0 1cm}]{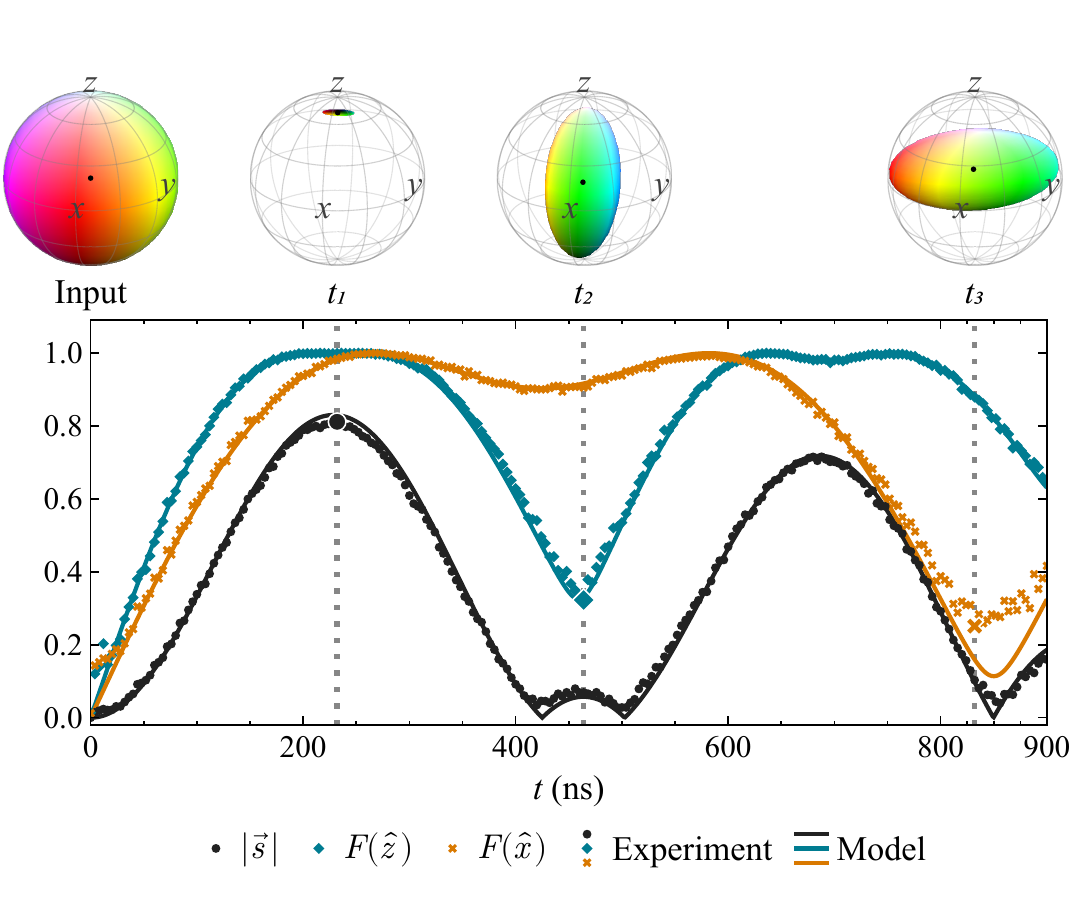}
    \caption{
    Time evolution of the quantities \(\abs{\s}\), \(F(\hat z)\), and \(F(\hat x)\) that enter the three-time quantum memory witness \(\wit\) in Eq.~\eqref{eq:triquantum-witness-definition} for \(\omega/g = 0.32\). The three times that minimize \(\wit\) are marked and the corresponding channels are visualized above within the Bloch ball.}
    \label{fig:SFF_bloch_032}
\end{figure}

\textit{Results.---}Figure~\ref{fig:SFF_bloch_032} displays $\abs{\s(t)}$, $F_t(\hat z)$, and $F_t(\hat x)$ versus time for \(\omega/g = 0.32\), together with the curves predicted by the model dynamics in Eq.~\eqref{eq:model-dynamics}. The three times \(t_{1} = 232\,\si{n s}\), \(t_2 = 464\,\si{n s}\), \(t_3 = 832\,\si{n s}\) that maximally violate the classical memory condition~\eqref{eq:triquantum-witness-condition} are marked. The corresponding channels \(\cpt_{1}\), \(\cpt_2\), \(\cpt_3\) are visualized as Bloch sphere images. For the given example we find \(\wit = -0.24 \pm0.03\), i.e., a strong signature of three-time quantum memory.

In order to show that the observed quantum memory is \emph{genuinely} three-time, we verify the existence of a two-time classical memory decomposition~\eqref{eq:biclassical-decomposition} for each pair of measured channels in a time window \(t \in \Tcal = (0,T)\) with \(T \geq t_3\).  
To this end, we numerically search for such decompositions via non-convex optimization. We consider two-time classical memory confirmed if we can reproduce the measured channels up to numerical accuracy (see SM \ref{sm::numericaloptimization} for details). 

\begin{figure}
    \centering
    \includegraphics[width=\linewidth]{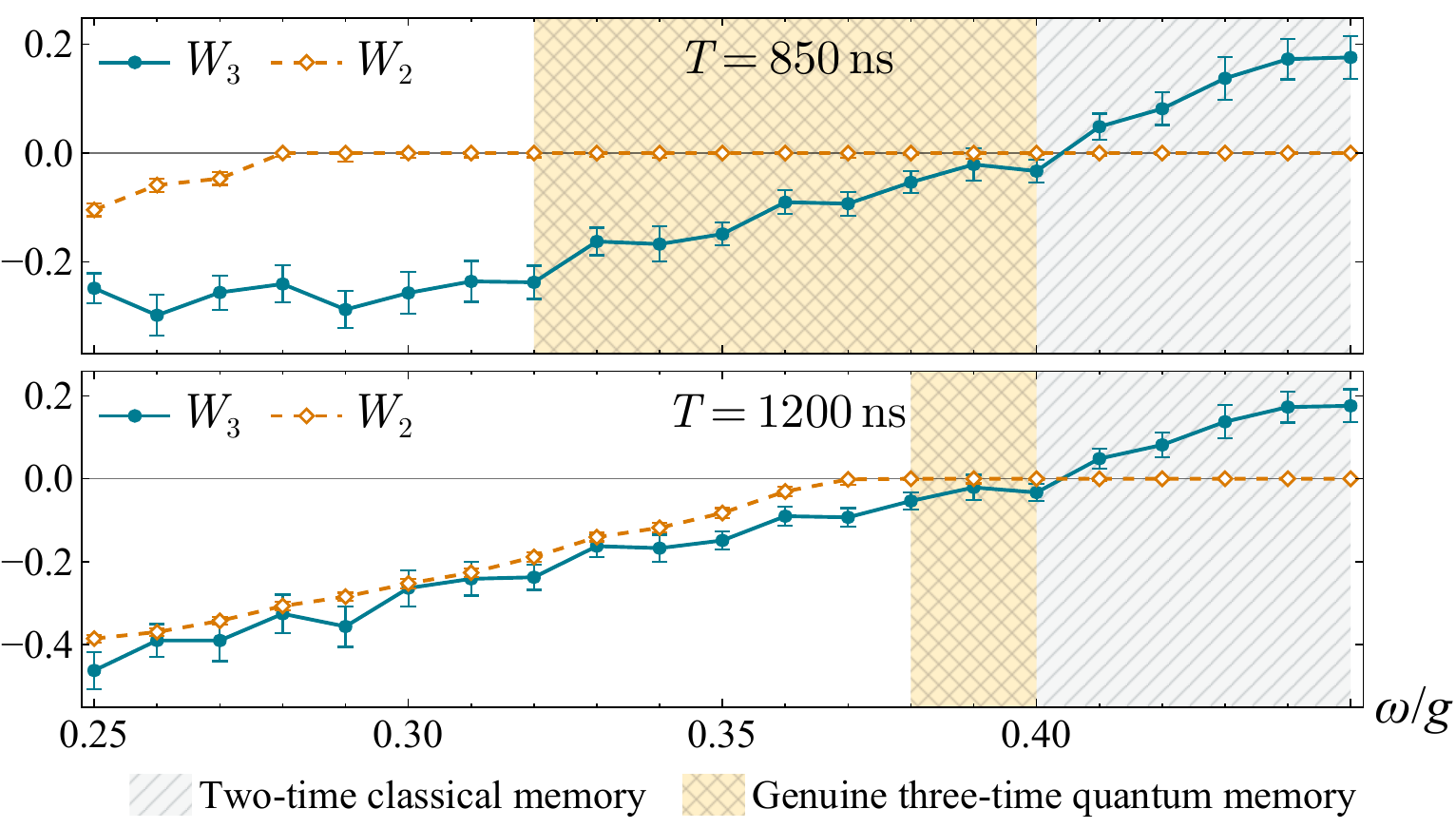}
    \caption{We plot the witness \(\wit\) versus \(\omega/g\) for two total observation times \(T \in \{850\,\si{n s}, 1200\,\si{n s}\}\) (upper and lower panels respectively). The error bars show the statistical uncertainty due to the channel tomography (see SM~\ref{sm::experimental-calibration}). We mark the parameter range for which a classical two-time decomposition can be numerically verified and for which the dynamical map exhibits genuine three-time quantum memory effects in the observed time window. For comparison we also plot the analytical two-time witness \(W_2\) from Ref.~\cite{backer_local_2024}.}
    \label{fig:witness-over-ratio}
\end{figure}

Importantly, the observability of quantum memory generally depends on the considered time domain \(\Tcal\). 
For example, quantum memory that is undetectable on the pairwise level in a certain time interval might show up in two-time witnesses if longer evolution times are considered. To illustrate this, we analyze the two- and three-time quantum memory of the measured dynamics in two time windows \(t \in (0, T)\) with \(T \in \{850\,\si{n s}, 1200\,\si{n s}\}\).
For driving strengths \(0.25 \leq \omega/g \leq 0.45\), we minimize the three-time witness \(\wit\) over all ordered triples \(t_1 < t_2 < t_3\) for the measured dynamics in the given time window. 
Then we numerically check if all pairs of maps in that window are classically decomposable. 
Additionally, we minimize a two-time quantum memory witness \(\BBS\) over all ordered pairs \(t_1 < t_2\) in the considered time interval. 
This two-time witness is defined by
\begin{align}
\label{eq:BBS}
    \BBS := C^\sharp[\chi_1] - C[\chi_2], 
\end{align}
where \(C\) and \(C^\sharp\) are the concurrence and the concurrence of assistance~\cite{laustsenLocalVsJoint2002a,divincenzoEntanglementAssistance1999}, respectively, and \(\chi_i\) is the Choi state of the channel \(\cpt_i\) (see Ref.~\cite{backer_local_2024} for details). 
The results are shown in Fig.~\ref{fig:witness-over-ratio}.

For small driving strengths \(\omega/g < 0.32\) where the non-Markovian amplitude damping [see Eq.~\eqref{eq:Hamiltonian-SA}] dominates, quantum memory is detectable at both the two- and three-time level. 
For strong driving \(\omega/g > 0.4\), no quantum memory can be verified by the witnesses  \(W_2\) and  \(W_3\). 
For an observation time \(T = 850\,\si{n s}\), a large window of intermediate driving strengths exists where we find genuine three-time quantum memory effects.
In contrast, for the longer observation of \(1200\,\si{n s}\), most ratios \(\omega/g\) that show three-time quantum memory also exhibit two-time quantum memory, and only a small window exists where we can confirm genuine three-time quantum memory. 
This is an expected behavior since the dynamics generated by the model Hamiltonian is (quasi)periodic, leading to (theoretically) perfect revivals of the input states at some point. 
In the experiment, the qubits are subject to additional noise processes that affect both the two- and three-time quantum memory signatures but do not fully wash out either.
However, in SM~\ref{sm::eternal-biclassicality} we show that dynamics which have classical two-time memory for arbitrary time windows while still exposing three-time quantum memory exist. 

\textit{Conclusions.---}In this Letter, we have demonstrated a strict hierarchy of multi-time quantum memory effects on the level of dynamical maps alone.
A quantum dynamical map whose memory effects can be explained classically for each pair of times can be incompatible with a classical decomposition if maps at three different times are considered jointly. 
We have developed a witness for three-time quantum memory and provided a prototypical discrete example for genuine three-time quantum memory.
The phenomenon is generic and also shows up in continuous dynamics, as we have experimentally demonstrated on a superconducting-qubit platform. 

Our framework naturally extends beyond three times (see SM~\ref{sm::n-classicality}). While the three-time witness \(\wit\) offers a particularly transparent analytical criterion, witnesses that rule out classical memory models based on semidefinite programming may provide a more direct numerical route to probing higher levels of the hierarchy (see SM~\ref{sm::SDP}). Establishing \emph{genuine} \(k\)-time quantum memory, however, becomes increasingly demanding: one must both detect quantum memory in the \(k\)-time dynamics and demonstrate that every \((k-1)\)-time subdynamics admits a classical-memory decomposition. This challenge is similar to that of characterizing genuinely multipartite entanglement~\cite{guhneEntanglementDetection2009,jungnitschTamingMultiparticleEntanglement2011,miklinMultiparticleEntanglementEmergent2016}.
Similarly, extensions of the witnesses to higher-dimensional quantum systems are, in principle, straightforward (see SM~\ref{sm::dimensiond}) but will require more dedicated research regarding their practical applicability. 

The central advantage of our approach is the experimental accessibility. Dynamical maps are routinely measured objects in quantum experiments~\cite{hashimPracticalIntroductionBenchmarking2025a,mohseniQuantumprocessTomographyResource2008,samachLindbladTomographySuperconducting2022}, and our results show that they carry information about memory beyond what pairwise analysis reveals alone. How much of this information can be ultimately unlocked from the full continuous dynamics $\cpt_t$ is a challenging open question. Answering this could turn standard tomographic data into a powerful probe that uses the accessible system dynamics to reveal complex quantum memory effects in an environment that is itself never measured directly.

\vspace{1em}
{\textit{Acknowledgments.---}}We thank Charlotte Bäcker for carefully reading the manuscript. KB and IP acknowledge support by the National Science Foundation under Grant No. 2239498, by NASA under Grant No. 80NSSC25K7051 and by the Alfred P. Sloan Foundation under Grant No. G-2023-21102. SSB, QC, OL and KM acknowledge support from National Science Foundation under Grant No.~2408932. 
VL acknowledges support by the Deutsche Forschungsgemeinschaft (DFG, German Research Foundation) via the Research Unit FOR 5688 (Project No. 521530974). The authors declare the use of generative AI (GPT-6 Astra, Claude Opus 5, Claude Fable 5). The following tasks were partially aided by AI tools under full human supervision: literature search, code generation, refinements of analytical proofs, and proofreading. 

\bibliography{references}

\appendix

\section*{Appendix}
\section{Proof of the Three-Time Quantum Memory Witness}
\label{app:proof}

We have to show that \(\wit \geq 0\) for any three-time classical-memory decomposition defined in Eq.~\eqref{eq:triclassical-decomposition}.
Summing over index \(j\) in Eq.~\eqref{eq:triclassical-decomposition} the maps \(\cpt_2\) and \(\cpt_3\) admit a form
\begin{align}
\label{eq:lambda-upsilon}
    \cpt_2[\rho] = \sum_i \Lambda_i[K_i \rho K_i^\dagger], \quad  \cpt_3[\rho] = \sum_i \Upsilon_i[K_i\rho K_i^\dagger],
\end{align}
with sets of CPTP maps \(\{\Lambda_i\}\) and \(\{\Upsilon_i\}\).
The witness is based on an incompatibility relation between the \(\{\Lambda_i\}\) and \(\{\Upsilon_i\}\) for shared \(\{K_i\}\).

\textit{Bound for \(\abs{\s_1}\)---}We define and decompose the operators
\begin{align}
\label{eq:ab-decomposition}
    E_i = K_i^\dagger K_i = a_i \id + \vb b_i \cdot \vec \sigma, && \tilde E_i = K_i K_i^\dagger = \tilde a_i \id + \tilde{\vb b}_i \cdot \vec \sigma.
\end{align}
\(E_i\) and \(\tilde E_i\) have the same spectrum.
Therefore, for the trace part we have \(a_i = \tilde a_i\).
Furthermore, \(\abs{\vb b_i} = \abs{\tilde{\vb b}_i}\), since \(\mathrm{tr}[(K_i^\dagger K_i)^2] = \mathrm{tr}[(K_i K_i^\dagger)^2] = 2(a_i^2 + \abs{\vb b_i}^2) = 2(\tilde a_i^2 + \abs{\tilde{\vb b}_i}^2)\).
Accordingly, since \(\cpt_1[\id] = \sum_i \tilde E_i\) we find for the vector \(\s_1\) defined in Eq.~\eqref{eq:s-displacement}
\begin{align}
\label{eq:abs-t}
    \abs{\s_1} = \abs{\sum_i \tilde{\vb b}_i} \leq \sum_i \abs{\tilde{\vb b}_i} = \sum_i \abs{{\vb b}_i}.   
\end{align}

\textit{Bound for \(F(\hat n)\)---}We define the flagged states 
\begin{align}
    \Omega_\pm^{\hat n} = \sum_i \ketbra{i}{i} \otimes \Kcal_i[\rho_\pm^{\hat n }], \quad \Kcal_i[\rho] := K_i \rho K_i^\dagger.
\end{align}
For the fidelity between these two states we find
\begin{align}
\label{eq:fidelity-sum}
     F&(\Omega_+^{\hat n}, \Omega_-^{\hat n}) = \norm{\sum_i \ketbra{i}{i} \otimes \sqrt{\Kcal_i[\rho_+^{\hat n}]}\sqrt{\Kcal_i[\rho_-^{\hat n}]}}_\text{tr} \\
     &= \sum_i \norm{\sqrt{\Kcal_i[\rho_+^{\hat n}]}\sqrt{\Kcal_i[\rho_-^{\hat n}]}}_\text{tr} =\sum_i F(\Kcal_i[\rho_+^{\hat n}], \Kcal_i[\rho_-^{\hat n}]) .\notag
\end{align}
Fidelities can only increase under the application of a CPTP map \(\Xi\) (states cannot become more distinguishable through a quantum channel). We define \(
    \Xi[\Omega] = \sum_i \Lambda_i[\bra{i} \Omega \ket{i}]\).
Eq.~\eqref{eq:fidelity-sum} can then be bounded by
\begin{align}
    F\left(\Omega_+^{\hat n}, \Omega_-^{\hat n}\right) \leq F\left(\Xi[\Omega_+^{\hat n}], \Xi[\Omega_-^{\hat n}]\right) = F_2(\hat n ),
\end{align}
with \(F_2(\hat n )\) defined in Eq.~\eqref{eq:Fk}. 
Since we consider pure input states \(\rho_\pm\) [see Eq.~\eqref{eq:antipodal-states}], the right-hand side of Eq.~\eqref{eq:fidelity-sum} can be written as
\begin{align}
\label{eq:fidelity-to-cross}
    \sum_i &F(\Kcal_i[\rho_+^{\hat n}], \Kcal_i[\rho_-^{\hat n}]) = \sum_i \abs{\bra{+_{\hat n}}K_i^\dagger  K_i \ket{-_{\hat n}}} \notag \\
    &= \sum_i \abs{\bra{+_{\hat n}} E_i\ket{-_{\hat n}}}  = \sum_i \abs{\vb b_i \cross \hat n},
\end{align}
where the \(\vb b_i\) come from Eq.~\eqref{eq:ab-decomposition}. The last equality can be understood by choosing w.l.o.g.\ \(\hat n = \hat z\), such that \(\ket{+_{\hat n}} = \ket{0}\), \(\ket{-_{\hat n}} = \ket{1}\) and \(
    \bra{0}\sigma_x\ket{1} = 1,\,  \bra{0}\sigma_y\ket{1} = -i,\, \bra{0}\sigma_z\ket{1} = 0\),
so \(    \abs{\bra{+_{\hat n}} E\ket{-_{\hat n}}} = \abs{\bra{0} \vb b \vec\cdot\sigma \ket{1}} = \abs{b_x - i b_y} = \abs{\vb b \cross \hat n}\).

Thus, with Eqns.~(\ref{eq:fidelity-sum}, \ref{eq:fidelity-to-cross}) and the same reasoning for the map \(\cpt_3\) with input state direction \(\hat m\) we have
\begin{align}
\label{eq:fidelity-bounds}
    \sum_i \abs{\vb b_i \cross \hat n} \leq F_2(\hat n), && \sum_i \abs{\vb b_i \cross \hat m} \leq F_3(\hat m).
\end{align}

\textit{Inequality---}In a last step we use  \(\abs{\vb b_i} \abs{\hat n \cross \hat m} \leq \abs{\vb b_i \cross \hat n} + \abs{\vb b_i \cross \hat m}\) (see SM~\ref{sm::cross-product}).
Combined with Eq.~\eqref{eq:abs-t} and Eq.~\eqref{eq:fidelity-bounds} this leads to 
\begin{gather}
    \abs{\s_1} \abs{\hat{n} \cross \hat m}  \leq \sum_i 
    \abs{\vb b_i} \abs{\hat n \cross \hat m} \\
    \leq \sum_i \abs{\vb b_i \cross \hat n} + \sum_i \abs{\vb b_i \cross \hat m} \leq  F_2(\hat n) + F_3(\hat m), \notag
\end{gather}
valid for any three-time classical memory dynamics. Thus, \(\wit\) in Eq.~\eqref{eq:triquantum-witness-definition} is a three-time quantum memory witness. \hfill \(\blacksquare\)

\section{Discrete Example}\label{sm::discreteexample}
We now provide a prototypical example of three maps that have two-time classical memory but three-time quantum memory.
The first map \(\cpt_1\) is an amplitude damping channel with canonical Kraus operators
\begin{align}
\label{eq:damping}
A_1 = \sqrt{\gamma} \sigma_- , && A_2 = \sqrt{1-\gamma} \sigma_+\sigma_-  + \sigma_-\sigma_+ , && 0<\gamma\leq 1.    
\end{align}
The second map \(\cpt_2\) is a dephasing channel around the \(z\)-axis with Kraus representation

\begin{align}
\label{eq:z-dephasing}
    Z_1 = \sqrt{p} \sigma_z, && Z_2 = \sqrt{1-p} \id, && 0\leq p \leq \frac{1}{2}.
\end{align}
The third map \(\cpt_3\) is a dephasing channel around the \(x\)-axis with the same dephasing strength \(p\) as \(\cpt_2\) and the Kraus representation
\begin{align}
\label{eq:x-dephasing}
    X_1 = \sqrt{p} \sigma_x, && X_2 = \sqrt{1-p} \id, && 0\leq p \leq \frac{1}{2}.
\end{align}
Since \(\cpt_2\) is a unital qubit map and therefore of mixed-unitary form, the pair \((\cpt_2, \cpt_3)\) is always two-time classical~\cite{backer_local_2024,backerQuantumMemoryPrecludes2026a}. 
The two other pairs \((\cpt_1, \cpt_2)\) and \((\cpt_1, \cpt_3)\) can be explained by classical memory if the dephasing strength \(p\) is sufficiently large. 
We explicitly construct classical memory decompositions of the form in Eq.~\eqref{eq:biclassical-decomposition} in App.~\ref{sm::biclassical-decompositions}, showing that the dynamics has two-time classical memory whenever 
\begin{align}
\label{eq:critical-dephasing}
    p \geq \frac{1}{2}(1- \sqrt{1-\gamma}).
\end{align}
If the dephasing \(p\) is weaker than the critical value in Eq.~\eqref{eq:critical-dephasing}, the dynamics has two-time quantum memory as can be demonstrated with the witness in Eq.~\eqref{eq:BBS}.

We now apply the three-time quantum memory witness in Eq.~\eqref{eq:triquantum-witness-definition} to this dynamics and show that it becomes negative for any choice of \(p\) and \(\gamma\).
Since the dephasing maps \(\cpt_2\) and \(\cpt_3\) preserve the antipodal eigenstates of \(\sigma_z\) and \(\sigma_x\), respectively, we choose \(\hat n = (0, 0, 1)^\top =: \hat z \) and \(\hat m = (1,0,0)^\top =: \hat x\) and get
\begin{align}
    F_2(\hat{z}) = F_3(\hat{x}) = 0.
\end{align}
This leads to the witness value
\begin{align}
    \wit = F_2(\hat{z}) + F_3(\hat{x}) -\abs{\s_1} = -\gamma.
\end{align}
Accordingly, the dynamics has three-time quantum memory for any non-trivial damping \(\gamma>0\) at time \(t_1\).

%%%%%%%%%%%%%%%%%%%%%%%

\section{The Two-time Classical Decompositions}
\label{sm::biclassical-decompositions}
\subsection{\(\cpt_1\) and \(\cpt_2\)}
We are looking for a decomposition of the form
\begin{align}
    \cpt_1 = \sum_i K_i \bullet K_i^\dagger, && \cpt_2 = \sum_i \Phi_i[K_i \bullet K_i^\dagger].
\end{align}
For the channels \(\cpt_1\) in Eq.~\eqref{eq:damping} and \(\cpt_2\) in Eq.~\eqref{eq:z-dephasing} the \(K_i\) and \(\Phi_i\) are given explicitly by
\begin{align}
    K_1 = A_1, && K_2 = A_2,
\end{align}
with \(A_i\) as in Eq.~\eqref{eq:damping}, and the channels \(\Phi_i\) in Kraus representation
\begin{align}
    \Phi_1:& \quad G_{1|1} = \sigma_+, && G_{2|1} = \sigma_+\sigma_- \\
    \Phi_2:& \quad G_{1|2} = \sqrt{f} \sigma_z, && G_{2|2} = \sqrt{1-f}\id, 
\end{align}
with
\begin{align}
\label{eq:f}
    f = {\frac{\sqrt{1-\gamma} + 2p - 1}{2\sqrt{1-\gamma}}}.
\end{align}
Thus, the classical decomposition is realized by measuring the standard Kraus operators of the amplitude damping in the first step.
If the state jumps (\(K_1 = A_1\)), a full reverse amplitude damping towards \(\ket{1}\) is implemented in the second step (\(\Phi_1\)). 
For outcome \(2\), the subsequent channel is a \(z\)-dephasing of strength \(f\).
This \(f\) has to be in the valid range \(0 \leq f \leq 1/2\), which yields a minimal dephasing condition for the parameter \(p\) in \(\cpt_2\), given the damping \(\gamma\) in \(\cpt_1\).
\begin{align}
\label{eq:critical-p}
    p \geq \frac{1}{2}(1- \sqrt{1-\gamma}).
\end{align}

The tightness of this bound follows from the quantum memory criterion in Eq.~\eqref{eq:BBS} which certifies quantum memory for any \(p\) below this bound.

\medskip

\subsection{\(\cpt_1\) and \(\cpt_3\)}
We are now looking for a decomposition of the form
\begin{align}
    \cpt_1 = \sum_i K_i \bullet K_i^\dagger, && \cpt_3 = \sum_i \Phi_i[K_i \bullet K_i^\dagger].
\end{align}
For the channels \(\cpt_1\) in Eq.~\eqref{eq:damping} and \(\cpt_3\) in Eq.~\eqref{eq:x-dephasing} the \(K_i\) and \(\Phi_i\) are given explicitly by
\begin{align}
    K_+ = \frac{A_1 + A_2}{\sqrt{2}}, && K_- = \frac{A_1 - A_2}{\sqrt{2}},
\end{align}
The channels \(\Phi_i\) are rotations around the \(y\)-axis whose direction depends on the outcome \(i\), followed by an \(x\)-dephasing
\begin{align}
    \Phi_\pm = \Dcal[R_\pm \bullet R_\pm^\dagger], 
\end{align}
with 
\begin{align}
    R_\pm = \exp[\pm i \phi \sigma_y /2], && \phi = - \arcsin\sqrt{\gamma}, 
\end{align}
and \(\Dcal\) given by the Kraus operators
\begin{align}
    G_1 = \sqrt{f} \sigma_x, && G_2 = \sqrt{1-f} \id, 
\end{align}
with the same \(f\) as in Eq.~\eqref{eq:f}, thus leading to the same critical dephasing \(p\) given in Eq.~\eqref{eq:critical-p}.

\medskip

\subsection{\(\cpt_2\) and \(\cpt_3\)}
The maps are unital qubit channels and, thus, a mixed unitary representation with classical memory exists~\cite{backer_local_2024,backerQuantumMemoryPrecludes2026a}.

\clearpage

\title{Supplemental Material}

\maketitle

\setcounter{section}{0} % or chapter
\renewcommand{\thesection}{\Roman{section}}
\renewcommand{\appendixname}{Supplement}

\counterwithout{equation}{section}

\setcounter{page}{1}
\renewcommand{\theequation}{S\arabic{equation}}
\setcounter{equation}{0}

\section{Experimental Calibration and Channel Reconstruction}
\label{sm::experimental-calibration}

In this section, we describe the experimental calibration steps to obtain the results in the main text. The detailed experimental parameters and setup can be in found Ref.~\cite{Qian2026}. 

We measure the readout confusion matrix $R$ by preparing and measuring the four computational basis states to obtain
\begin{equation}
R=
\begin{pmatrix}
0.98195 & 0.05415 & 0.04750 & 0.00285\\
0.00525 & 0.93455 & 0.00015 & 0.04495\\
0.01270 & 0.00065 & 0.94710 & 0.05680\\
0.00010 & 0.01065 & 0.00525 & 0.89540
\end{pmatrix}.
\label{eq:joint-response-matrix}
\end{equation}
Here, $R_{m,j}$ is the probability of measuring $\ket{m}$ after preparing $\ket{j}$. The columns label prepared states and the rows label measured states. The diagonal elements represent correct assignments, whereas the off-diagonal elements represent misidentification errors. The readout fidelity is $93.975\%$. When characterizing the channel, we  correct the raw measurement outcomes using constrained multinomial maximum-likelihood estimation~\cite{Maciejewski2020,Bravyi2021}. 

The coupling Hamiltonian $H_{SA}$ is generated by parametric modulation. This modulation also  shifts the transition frequencies of the two qubits. To determine the frequency shift, we prepare $\ket{00}$, apply the parametric modulation and a local drive on $S$ simultaneously. We sweep the local drive frequency and duration. We measure the state of two qubits and fit the two qubits' state probabilities to probabilities calculated by propagating the initial state according to the Hamiltonian:
\begin{align}
\frac{H_{\mathrm{fit}}}{\hbar}
={}&\frac{H_{SA}}{\hbar}
-\delta_d(n_S+n_A)
\nonumber\\
&+\frac{\Delta_{\mathrm{ex}}}{2}(n_S-n_A)
+\omega\sigma_x^S .
\label{eq:experimental-fit-Hamiltonian}
\end{align}
Here, $n_j=\ketbra{1}{1}_j$ is the excitation-number operator,
$\delta_d$ is the common detuning of the two single-excitation states from
the scanned local drive, and $\Delta_{\mathrm{ex}}$ is their residual
relative detuning in the rotating frame of the parametric modulation. To see
how these parameters determine the probe resonance, consider the diagonal
energies of the two single-excitation states:
\begin{align}
\frac{E_{10}}{\hbar}
&=
-\delta_d+\frac{\Delta_{\mathrm{ex}}}{2},\\
\frac{E_{01}}{\hbar}
&=
-\delta_d-\frac{\Delta_{\mathrm{ex}}}{2}.
\end{align}
The state $\ket{10}$ corresponds to an excitation of the probe $S$.
Therefore, the local drive is resonant with the probe when $\delta_d=\frac{\Delta_{\mathrm{ex}}}{2}$. We vary the local drive frequency and duration and use the measured two-qubit probability outcomes to find the probe qubit's resonance frequency when the parametric modulation is applied: $\omega_{S}/(2\pi) = 4.078\,\si{GHz}$. 

The time evolution in the presence of drive and exchange interaction generates a dynamical phase on the probe qubit. We measure the resulting dynamical phase using a Ramsey
exchange-echo sequence. We prepare a superposition state on $S$, we apply the exchange interaction for $t/2$, shift the phase of the parametric modulation by $\pi$ to reverse $g\rightarrow-g$, and apply the exchange interaction for another $t/2$. The second half reverses the population exchange. We then sweep the phase of the final Ramsey pulse to determine the accumulated dynamical phase. This phase is then compensated for using virtual frame rotations allowing reliable measurements in the three Pauli bases.

\subsection{Witness Uncertainty}
\label{sec:uncertainty}
A qubit state can be represented by its Bloch vector $\mathbf r$ as
\begin{equation}
\rho=\frac{1}{2}
\left(\id+\mathbf r\cdot\Vec{\sigma}\right).
\end{equation}
The reconstructed channel maps the input Bloch vector $\mathbf r_{\mathrm{in}}$ to the output Bloch
vector $\mathbf r_{\mathrm{out}}$ according to
\begin{equation}
\mathbf r_{\mathrm{out}}
=
A\mathbf r_{\mathrm{in}}+\s.
\end{equation}
Here, the $3\times3$ matrix $A$ describes the rotation and contraction
of the Bloch sphere, while $\s$ is the displacement vector defined in the
main text. For each of the six known input Bloch vectors, measurements in the three Pauli
bases determine the corresponding output Bloch vector. We fit all measured outcomes simultaneously to $\mathbf r_{\mathrm{out}}=A\mathbf r_{\mathrm{in}}+\s$ using
CPTP-constrained maximum-likelihood estimation. This fit determines the nine
elements of $A$ and the three components of $\s$. These fitted parameters are used to construct a $12\times12$ covariance matrix $\Sigma$. For each witness ingredient $X\in\{\abs{\s_1},F_2(\hat z),F_3(\hat x)\}$, we propagate this uncertainty as
\begin{equation}
\sigma_X^2
=
(\nabla X)^{\mathsf T}\Sigma(\nabla X),
\end{equation}
where $\nabla X$ describes how $X$ changes when the fitted parameters are varied. Since the three witness ingredients are measured independently, the one-standard-deviation uncertainty of the witness is
\begin{equation}
\sigma_{\wit}
=
\sqrt{
\sigma_{\abs{\s_1}}^2+
\sigma_{F_2(\hat z)}^2+
\sigma_{F_3(\hat x)}^2
}.
\end{equation}
This uncertainty is the error bar shown in Fig.~\ref{fig:witness-over-ratio}.

\section{Numerical Optimization of Two-time Classical-Memory Realizations}\label{sm::numericaloptimization}

Since we do not have an exact criterion to check if a pair of channels has a classical realization according to Eq.~\eqref{eq:biclassical-decomposition}, we use a numerical optimization algorithm that searches for the closest classical memory process that is compatible with the observed channels. Our procedure works by first computing a Kraus decomposition for $\cpt_1$
\begin{equation}
    \cpt_1[\rho]=\sum_{i=1}^{d^2} K_i\rho K_i^\dagger .
\end{equation}
Different Kraus decompositions are related by isometries like
\begin{align}
    \tilde K_i = \sum_j V_{ij} K_j,
\end{align}
where \(V\) lives on the complex Stiefel manifold \(\operatorname{St}(r,d^2)\), i.e.~$V^\dagger V=\id$, and $r\geq d^2$. This form ensures that the map $\cpt_1$ is represented exactly. In order to find a classical decomposition for map $\cpt_2$ we need to further find the conditional channels $\Phi_i$. These can likewise be parameterized through their Kraus decomposition
\begin{equation}
    \Phi_i[\rho]=\sum_{j=1}^{d^2} G_{j|i}\rho G^\dagger_{j|i},\qquad \sum_{j=1}^{d^2} G^{\dagger}_{j|i} G_{j|i} =\id
\end{equation}
By stacking the operators $G_{j|i}$ for fixed $i$ in a single matrix one again obtains an isometry $(G_{1|i},\ldots, G_{d^2|i})\in \mathrm{St}(d^3,d)$. 
We can then define a second map \(\tilde \Ecal_2\) through
\begin{align}
    \tilde \Ecal_2[\rho] = \sum_{i,j}  G_{j|i} \tilde K_i \rho  \tilde K_i^\dagger  G_{j|i}^\dagger,
\end{align}
which parametrizes all possible classical memory decompositions consistent with $\cpt_1$ through points on a product manifold
\begin{align}
    \mathcal{M} = \operatorname{St}(r, d^2) \times [\operatorname{St}(d^3,d)]^{\times r}. 
\end{align}
If we find a decomposition such that $\tilde \cpt_2=\cpt_2$, the pair of maps is classically realizable. 
Note that the maximum memory rank $r$ is bounded \cite{yu_quantum_2025}. %can be bounded by $r\leq (d^2-1)(2d^2+1)$.
The Stiefel manifold is a standard manifold for Riemannian optimization. 
As a cost function for a numerical optimization we use the fidelity \(F\) of the corresponding Choi states of $\tilde \cpt_2$ and $\cpt_2$. We use the quasi-Newton algorithm implemented in Manopt.jl~\cite{Bergmann_2022} which converges reliably for this problem. 
{We accept a pair of channels to be realizable with classical memory if the optimization reaches a numerical infidelity of \(1- F < 10^{-10}\)}.

%%%%%%%%%%%%%%%%%%%%%%%%%%

\section{Eternally Two-time Classical Memory Dynamics}
\label{sm::eternal-biclassicality}
As we have seen in the main text, the question of whether quantum memory is detectable can strongly depend on the time window the dynamical map is measured for. 
The comparison of the two different windows in Fig.~\ref{fig:witness-over-ratio} shows that quantum memory that is only detectable with a three-time witness in a short interval may become detectable with a two-time witness again if the measured data contains a longer observation. 
Obviously, it is trivial to engineer cases with three-time but no two-time quantum memory for example by using a time-dependent dissipation that increases strongly after the three-time witness becomes negative. 
However, also time-independent dissipation can lead to this behavior, often depending sensitively on the interplay of coupling ratios and decoherence rates. 

In this section we give an example that shows this behavior relatively strongly. The dynamics can numerically be tested to have two-time classical memory for all times but it nevertheless shows a strong three-time quantum memory signature.

\begin{figure}
    \centering
    \includegraphics[width=\linewidth]{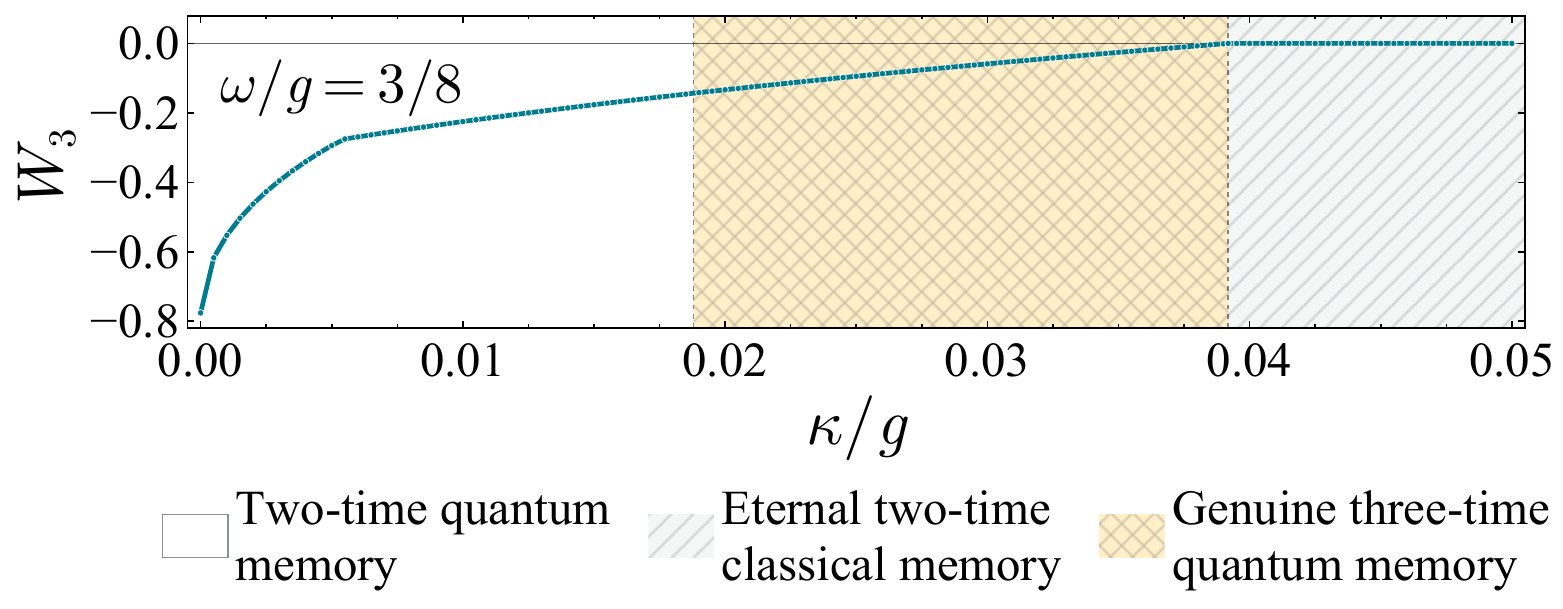}
    \caption{The three-time witness \(\wit\) evaluated at an optimal triple of times \(t_1<t_2<t_3\) in the time evolution with dissipator \(L\) and dissipation strength \(\kappa\). The two-time classical memory is numerically verified for all times in the single-hatched parameter range. In the yellow, double-hatched region, the dynamics shows genuine three-time quantum memory. The kink of \(\wit\) at \(\kappa/g \approx 0.005\) is not a numerical artifact. At this point the optimal choice of times \(t_1, t_2, t_3\) that minimizes \(\wit\) changes discontinuously.}
    \label{fig:eternal-memory}
\end{figure}

We use the dynamics described in the main text and add a Lindblad dissipator of the form 
\begin{align}
    L = \sigma_x^S \otimes \sigma_x^A
\end{align}
with rate \(\kappa\). 
Thus in the Markovian three-qubit embedding, the time evolution is described by the time-independent generator
\begin{align}
    \dot \rho_{SAB} = \Lcal \rho_{SAB} &= -\frac{i}{\hbar} [H_{SA}+H_{SB},\rho_{SAB}] \notag\\
    &+ \kappa \left( L \rho_{SAB} L^\dagger - \frac12 \{L^\dagger L, \rho_{SAB}\} \right),
\end{align}
with the anticommutator \(\{\bullet, \bullet\}\). The dynamics of the probe system \(S\) alone is then given by
\begin{align}
    \cpt_t[\rho_S] = \tr_{AB}\left[e^{\Lcal t}(\rho_S \otimes \ketbra{0}{0}_A \otimes \ketbra{0}{0}_B)\right].
\end{align}
We choose a rational driving strength \(\omega/g = 3/8\) which for \(\kappa=0\) leads to a cyclic dynamics with period \(\tau = 8\pi/g\). 
Every possible ordered pair of maps \((\cpt_1, \cpt_2)\) appears at least once within a total time of two periods, i.e.\ \(T = 2\tau = 16\pi/g\). 
For this cyclic case (\(\kappa=0\)), quantum memory can always be detected with two-time tests.
For \(\kappa > 0\), the cyclic behavior is broken. 
We numerically sample pairs densely within the time window \((0,T = 2\tau)\) and check their two-time quantum memory behavior as a function of the decoherence rate \(\kappa\). We do not need to consider longer times. Since the Lindblad operator \(L\) commutes with the Hamiltonian, it only adds additional decoherence in later periods. If all pairs are classical up to time \(T\), the dynamics will be two-time classical for any later time too.
The results are shown in Fig.~\ref{fig:eternal-memory}.
For large \(\kappa\) the decoherence dominates and no quantum memory can be detected. 
In an intermediate range there is a regime where the three-time quantum memory survives but the two-time snapshots are classical for arbitrary choices of time. Thus, this is the regime where the dynamics has genuine three-time quantum memory for arbitrary long observation times.

\section{\(\mathbf{\textit{k}}\)-time Quantum Memory Witness}\label{sm::n-classicality}
The concept can straightforwardly be extended to more than three maps.
Let $\mathcal{E}_1,\dots,\mathcal{E}_k$ be qubit CPTP maps at times
$t_1<\dots<t_k$ that admit a $k$-time classical-memory decomposition,
i.e., the generalization of Eq.~\eqref{eq:triclassical-decomposition} to $k-1$
ordered classical memory updates. Summing over all labels except the
first, every later map takes the form
\begin{equation}
  \mathcal{E}_l[\rho] = \sum_i \Lambda^{(l)}_i\!\left[K_i \rho K_i^\dagger\right],
  \qquad l = 2,\dots,k,
  \label{eq:shared_instrument}
\end{equation}
with CPTP maps $\{\Lambda^{(l)}_i\}$ and the same Kraus operators
$\{K_i\}$ of $\mathcal{E}_1$ for all $l$.
The argument leading to
Eq.~\eqref{eq:fidelity-bounds} therefore holds for each later map
separately. For any unit vectors $\hat{n}_2,\dots,\hat{n}_k$,
\begin{equation}
  \sum_i |\mathbf{b}_i \times \hat{n}_l| \le F_l(\hat{n}_l),
  \qquad l = 2,\dots,k,
  \label{eq:fidelity_bounds_k}
\end{equation}
where $K_i^\dagger K_i = a_i \id + \vb{b}_i\cdot\vec{\sigma}$
as in Eq.~\eqref{eq:ab-decomposition}.
We define the geometric constant
\begin{equation}
  c(\hat{n}_2,\dots,\hat{n}_k) := \min_{|\hat{u}|=1}\,
  \sum_{l=2}^{k} |\hat{u}\times\hat{n}_l| .
  \label{eq:c_def}
\end{equation}
Summing Eq.~\eqref{eq:fidelity_bounds_k} over $l$ and using
$|\vec{s}_1| \le \sum_i |\mathbf{b}_i|$ from Eq.~\eqref{eq:abs-t}, we find
\begin{equation}
  c\,|\vec{s}_1| \le c \sum_i |\mathbf{b}_i|
  \le \sum_i \sum_{l=2}^{k} |\mathbf{b}_i\times\hat{n}_l|
  \le \sum_{l=2}^{k} F_l(\hat{n}_l).
\end{equation}
Hence, for every $k$-tuple of channels with a $k$-time classical-memory
decomposition, the quantity
\begin{equation}
  W_k := \min_{\hat{n}_2,\dots,\hat{n}_k}
  \left[\, \sum_{l=2}^{k} F_l(\hat{n}_l)
  - c(\hat{n}_2,\dots,\hat{n}_k)\,|\vec{s}_1| \right]
  \label{eq:Wk}
\end{equation}
satisfies $W_k \ge 0$, and an observed violation $W_k<0$ certifies
quantum memory.

\section{SDP Witnesses}
\label{sm::SDP}
Two-time quantum memory can be witnessed with the help of semidefinite programs (SDP)~\cite{yu_quantum_2025}. 
An extension to the multi-time case is directly possible.

Let us start with a slightly modified form of the two-time SDP witness from Ref.~\cite{yu_quantum_2025}.
The Choi state of a CPTP map \(\cpt\) on a \(d\)-dimensional quantum system is defined by
\begin{align}
    J(\cpt) = (\id\otimes\cpt)[\ketbra{\Omega}{\Omega}],\qquad \ket{\Omega} = \sum_{k =0}^{d-1} \ket{kk}.
\end{align}
If two quantum channels \(\cpt_1\) and \(\cpt_2\) with Choi states \(J(\cpt_1)\) and \(J(\cpt_2)\), respectively, have a classical-memory decomposition as in Eq.~\eqref{eq:biclassical-decomposition}, then there exists a separable process tensor \(X_{\text{cl}}\) of the form
\begin{align}
\label{eq:classical-X}
    X^{A'BB'C}_{\text{cl}} = \sum_i J(\Kcal_i)^{A'B} \otimes J(\Phi_i)^{B'C},
\end{align}
with \(\Kcal_i[\bullet] = K_i \bullet K_i^\dagger\). This \(X_\text{cl}\) reproduces the two maps \(\cpt_1\) and \(\cpt_2\) as its marginals~\cite{yu_quantum_2025}:
\begin{gather}
    \tr_{C} X_{\text{cl}} = J(\cpt_1)^{A'B} \otimes \id^{B'}, \\
    \bra{\Omega_{BB'}} X_{\text{cl}} \ket{\Omega_{BB'}} = J(\cpt_2)^{A'C}.
\end{gather}
Any process tensor of this classical form~\eqref{eq:classical-X} has by construction a positive partial transpose (PPT) with respect to the cut \(A'B|B'C\).
We can use this as a constraint in the following SDP:
\begin{align}
   W_\text{SDP} =  -\min_{X} ||\bra{\Omega_{BB'}} &X \ket{\Omega_{BB'}} - J(\cpt_2)||  \\
\text{subject to } &X \succeq 0, \\
&X^{\top_{B'C}} \succeq 0, \label{eq:sdp-PT}\\
&\tr_{C} X = J(\cpt_1)^{A'B} \otimes \id^{B'}. \label{eq:sdp-marginal}
\end{align}
This SDP optimizes a positive-semidefinite \(X\) such that it reproduces the second channel \(\cpt_2\) as its marginal.
Line~\eqref{eq:sdp-marginal} guarantees that \(X\) also reproduces channel \(\cpt_1\) as its first marginal. 
The other constraint~\eqref{eq:sdp-PT} is the PPT requirement from above. 
If \(W_\text{SDP} < 0\), then there is no PPT process tensor \(X\) that reproduces both maps \(\cpt_1\) and \(\cpt_2\), and, accordingly a classical-memory decomposition cannot exist. 
Therefore, \(W_\text{SDP}\) is a witness for two-time quantum memory.

Adding another time step is straightforward. 
The dynamics now emerges from a process tensor \(X^{A'BB'CC'D}\).
If there is a classical-memory decomposition, it can be written as the tensor
\begin{align}
\label{eq:3-tensor}
    X^{A'BB'CC'D} = \sum_{ij} J(\Kcal^{A'B}_i) \otimes J(\Gcal^{B'C}_{j|i}) \otimes J(\Phi^{C'D}_{ij}),
\end{align}
with \(\Gcal_{j|i}[\bullet] = G_{j|i} \bullet G_{j|i}^\dagger\).
Therefore, this process tensor must have a positive partial transpose with respect to the three cuts \(A'B|B'CC'D\), \(A'BB'C|C'D\), and \(A'BC'D|B'C\)~\cite{taranto_2024,giarmatziWitnessingQuantumMemory2021,whiteWhatCanUnitary2025a}. 
Thus, in analogy to the SDP above we now have
\begin{align}
   W_\text{SDP} \,=\,  -&\min_{X} ||\bra{\Omega_{BB'}\Omega_{CC'}} X \ket{\Omega_{BB'}\Omega_{CC'}} - J(\cpt_3)||  \notag \\
\text{subject to } &X \succeq 0, \\
&X^{\top_{A'B}} \succeq 0, \quad X^{\top_{B'C}} \succeq 0, \quad X^{\top_{C'D}} \succeq 0,\label{eq:sdp-multi-PT}\\
&\tr_{CD} X = J(\cpt_1)^{A'B} \otimes \id^{B'} \otimes \id^{C'},\\
&\tr_{D} \bra{\Omega_{BB'}}X\ket{\Omega_{BB'}} = J(\cpt_2)^{A'C} \otimes \id^{C'}.
\end{align}
The last two constraints ensure that \(X\) reproduces the first and second marginal channels \(\cpt_1\) and \(\cpt_2\), respectively, and line~\eqref{eq:sdp-multi-PT} describes the PPT constraints.
If we find \(W_\text{SDP} < 0\), there is no compatible \(X\) of the form~\eqref{eq:3-tensor} and the dynamics cannot have a classical-memory decomposition.

Higher-order multi-time witnesses can be constructed in the very same way by adding the respective marginal constraints and PPT conditions along all cuts.

\section{Extending the Three-time Quantum Memory Witness $\mathbf{\textit{W}_3}$ to Dimension $\mathbf{\textit{d}}$}
\label{sm::dimensiond}
The analytical three-time witness \(\wit\) can be extended to \(d\)-dimensional systems. Let \(\cpt_1,\cpt_2,\cpt_3\) be quantum channels on the same \(d\)-dimensional system, corresponding to \(t_1<t_2<t_3\). For simplicity we choose two mutually unbiased orthonormal bases [corresponding to \(\abs{\hat n \cross \hat m} = 1\) in the qubit witness in Eq.~\eqref{eq:triquantum-witness-definition}],
\begin{align}
\mathcal B=\{|b_m\rangle\}_{m=1}^d,\quad
\mathcal C=\{|c_m\rangle\}_{m=1}^d,
\quad |\langle b_m|c_n\rangle|^2=\frac1d.
\end{align}
We define
\begin{gather}
\label{eq:SFF-d}
    \Scal_1  = \frac12\norm{\mathcal E_1(\id)-\id}_\text{tr} , \\
    F_2(\Bcal) = \sum_{m<n} F\left(\cpt_2[\ketbra{b_m}{b_m}], \cpt_2[\ketbra{b_n}{b_n}] \right), \notag\\
    F_3(\Ccal) = \sum_{m<n} F\left(\cpt_3[\ketbra{c_m}{c_m}], \cpt_3[\ketbra{c_n}{c_n}] \right). \notag
\end{gather}
The following quantity is a three-time quantum memory witness for a \(d\)-dimensional quantum system
\begin{align}
\label{eq:triquantum-witness-d}
    W_3^{(d)} = F_2(\Bcal) + F_3(\Ccal) - \Scal_1.
\end{align}

The proof is similar to the qubit case. We use again that the maps \(\cpt_2\) and \(\cpt_3\) admit a form
\begin{align}
\label{eq:lambda-upsilon-d}
    \cpt_2[\rho] = \sum_i \Lambda_i[K_i \rho K_i^\dagger], \quad  \cpt_3[\rho] = \sum_i \Upsilon_i[K_i\rho K_i^\dagger],
\end{align}
with shared \(\{K_i\}\).
We write
\begin{align}
\label{eq:Ei-decomp}
    E_i = K_i^\dagger K_i = a_i \id + X_i, \quad a_i = \frac{\tr E_i}{d}, \quad \tr X_i =0,
\end{align}
which satisfy \(\sum_i E_i = \id\) and therefore \(\sum_i a_i = 1\). 
Each \(K_i\) admits a polar decomposition \(K_i = U_i \sqrt{E_i}\) with unitary \(U_i\). Therefore,
\begin{align}
    \cpt_1[\id] - \id = \sum_i U_i E_i U_i^\dagger - \id = \sum_i U_i X_i U_i^\dagger.
\end{align}
The triangle inequality and the unitary invariance of the trace norm give
\begin{align}
    \Scal_1 \leq \frac12 \sum_i \norm{X_i}_\text{tr}.
\end{align}
So the nonunitality of \(\cpt_1\) is bounded by the norm of the traceless part of the elements \(E_i\).

We can bound these parts further by using their off-diagonal entries with respect to the two bases \(\Bcal\) and \(\Ccal\).
We define \(\Dcal_\Bcal\) and \(\Dcal_\Ccal\) as the fully dephasing channels with respect to basis \(\Bcal\) and \(\Ccal\), respectively.
Due to mutual unbiasedness their concatenation leads to 
\begin{align}
    \Dcal_\Bcal \circ \Dcal_\Ccal[X] = \frac{\tr X}{d}\id.
\end{align}
For a traceless \(X\) we can then write
\begin{align}
    X = (X - \Dcal_\Bcal[X]) + \Dcal_\Bcal[X-\Dcal_\Ccal[X]].
\end{align}
The dephasing map cannot increase the trace norm of a Hermitian operator. Thus, we find
\begin{align}
    \norm{X}_\text{tr} \leq \norm{X - \Dcal_\Bcal[X]}_\text{tr} + \norm{X - \Dcal_\Ccal[X]}_\text{tr}. 
\end{align}
Each term contains only off-diagonal entries with respect to basis \(\Bcal\) or \(\Ccal\), respectively.
Each Hermitian pair of entries \(x_{mn}\), \(x_{nm}=x_{mn}^*\) has trace norm \(2\abs{x_{mn}}\). 
Thus decomposing the operators into these pairs and using the triangle inequality we get 
\begin{align}
\label{eq:S-bound}
    \frac12 \norm{X}_\text{tr} \leq \sum_{m<n} \abs{\bra{b_m} X \ket{b_n}} + \sum_{m<n} \abs{\bra{c_m} X \ket{c_n}}. 
\end{align}
We now apply this to each \(X_i\) and make use of the fact that their off-diagonal entries are equal to those of \(E_i\) [see Eq.~\eqref{eq:Ei-decomp}] which yields
\begin{align}
\label{eq:Scal-bound}
    \Scal_1 \leq \sum_{m<n} \sum_i  \abs{\bra{b_m} E_i \ket{b_n}} + \sum_{m<n} \sum_i  \abs{\bra{c_m} E_i \ket{c_n}}. 
\end{align}

Now we derive bounds for the quantities \(F_2\) and \(F_3\).
As in the qubit case we define flagged states for two pure input states \(\ket u\) and \(\ket v\) after the application of the Kraus channel:
\begin{align}
    \Omega_u = \sum_i \ketbra{i}{i} \otimes \Kcal_i[\ketbra{u}{u}], && \Omega_v = \sum_i \ketbra{i}{i} \otimes \Kcal_i[\ketbra{v}{v}],
\end{align}
where \(\Kcal_i[\bullet] = K_i \bullet K_i^\dagger\).
Their fidelity is 
\begin{align}
\label{eq:fidelity-d}
    F(\Omega_u, \Omega_v) = \sum_i \abs{\bra{v}K_i^\dagger K_i \ket{u}} =  \sum_i \abs{\bra{v}E_i\ket{u}}
\end{align}
 We again use the CPTP map 
 \begin{align}
     \Xi[\Omega] = \sum_i \Lambda_i[\bra{i} \Omega \ket{i}],
 \end{align}
 with the \(\Lambda_i\) from Eq.~\eqref{eq:lambda-upsilon-d}.
The fidelity in Eq.~\eqref{eq:fidelity-d} cannot decrease under the  application of a CPTP map. Thus, we find the bound
\begin{align}
     \sum_i \abs{\bra{v}E_i\ket{u}}\leq F\big(\cpt_2[\ketbra{u}{u}],\cpt_2[\ketbra{v}{v}]\big).
\end{align}
Using pairs \(\ket{b_m}, \ket{b_n}\) and summing over them we get
\begin{align}
\label{eq:F-bound-d}
    &\sum_{m<n}\sum_i \abs{\bra{b_n}E_i\ket{b_m}} \notag \\
    &\leq \sum_{m<n} F\big(\cpt_2[\ketbra{b_m}{b_m}],\cpt_2[\ketbra{b_n}{b_n}]\big) = F_2(\Bcal),
\end{align}
with the \(F_2(\Bcal)\) from Eq.~\eqref{eq:SFF-d}.
The same reasoning leads to a bound for \(F_3(\Ccal)\).
Plugging Eq.~\eqref{eq:F-bound-d} into Eq.~\eqref{eq:Scal-bound} we arrive at
\begin{align}
    \Scal_1 \leq F_2(\Bcal) + F_3(\Ccal),
\end{align}
which holds for any dynamics with a classical memory decomposition as in Eq.~\eqref{eq:triclassical-decomposition}.
Thus, the quantity \(\wit^{(d)}\) in Eq.~\eqref{eq:triquantum-witness-d} is a three-time quantum memory witness.

\section{Cross Product Inequality}
\label{sm::cross-product}
We want to show that 
\begin{align}
        \abs{\vb b} \abs{\hat n \cross \hat m} \leq \abs{\vb b \cross \hat n} + \abs{\vb b \cross \hat m}.
\end{align}
We define the unit vector
\begin{align}
    \hat u = \frac{\vb b}{\abs{\vb b}}.
\end{align}
The unit vectors \(\hat n\) and \(\hat m\) can then be decomposed into a part parallel and perpendicular to \(\hat u\):
\begin{align}
    \hat n = (\hat n \cdot \hat u) \hat u + [\hat n - (\hat n\cdot \hat u)\hat u] = \vec n_\parallel + \vec n_\perp,
\end{align}
and equivalently for \(\hat m = \vec m_\parallel + \vec m_\perp\).
We now write
\begin{align}
    \hat n \cross \hat m = \vec n_\parallel \cross \vec m_\parallel +\vec n_\perp \cross \vec m_\perp + \vec n_\parallel \cross \vec m_\perp +\vec n_\perp \cross \vec m_\parallel .
\end{align}
The first term on the rhs is zero since \(\vec n_\parallel \parallel \vec m_\parallel\). The second term is orthogonal to the last two. Squaring both sides and using the triangle inequality we obtain
\begin{align}
    \abs{\hat n \cross \hat m}^2 &=  \abs{\vec n_\perp \cross \vec m_\perp}^2 + \abs{\vec n_\parallel \cross \vec m_\perp + \vec n_\perp \cross \vec m_\parallel}^2 \notag \\
    & \leq \abs{\vec n_\perp \cross \vec m_\perp}^2 + \left(\abs{\vec n_\parallel \cross \vec m_\perp} + \abs{\vec n_\perp \cross \vec m_\parallel}  \right)^2.
\end{align}
We define
\begin{align}
    x = \abs{\vec n_\perp}, && y = \abs{\vec m_\perp}.
\end{align}
Since \(\hat n\) and \(\hat m\) are unit vectors, we have
\begin{align}
    \abs{\vec n_\parallel} = \sqrt{1-x^2}, && \abs{\vec m_\parallel} = \sqrt{1-y^2}.
\end{align}

Now we can write
\begin{align}
     \abs{\hat n \cross \hat m}^2 &\leq x^2 y^2 + \left(y\sqrt{1-x^2} + x\sqrt{1-y^2}\right)^2 \notag \\
     & = (x + y)^2 - x^2y^2 - 2xy\left(1 - \sqrt{1 -x^2}\sqrt{1 -y^2} \right).
\end{align}
Since \(0 \leq x,y \leq 1\) we can drop the last two terms and arrive at
\begin{align}
    \abs{\hat n \cross \hat m} \leq x + y = \abs{\vec n_\perp} + \abs{\vec m_\perp} = \abs{\hat u \cross \hat n} + \abs{\hat u \cross \hat m}.    
\end{align}
Multiplying both sides by \(\abs{\vb b}\) we get the desired relation
\begin{align}
     \abs{\vb b} \abs{\hat n \cross \hat m} \leq \abs{\vb b \cross \hat n} + \abs{\vb b \cross \hat m}.
\end{align}

\end{document}